 \documentclass[final,5p,times,twocolumn,authoryear]{elsarticle}

\usepackage{amssymb}
\usepackage{subfigure}
\usepackage{hyperref}
\journal{Radiation measurements}

\begin{document}

\begin{frontmatter}

\title{Luminescence of photochromic centers in calcium fluoride crystals doped with Lu$^{3+}$ ions}

\author[igc,isu]{R. Shendrik}
\ead{r.shendrik@gmail.com}
\author[igc]{A.~S.~Myasnikova}
\author[igc]{T.~Yu.~Sizova}
\author[igc,isu]{E.~A.~Radzhabov}
\address[igc]{Vinogradov Institute of geochemistry SB RAS, Favorskogo 1a, Irkutsk,Russia, 664033}
\address[isu]{Physics department of Irkutsk state university, Gagarina blvd 20, Irkutsk, Russia, 664003}

\begin{abstract}
We report data on the luminescence spectra associated with photochromic centers in X-ray irradiated calcium fluoride crystals doped with Lu ions. Irradiation in low energy photochromic centers absorption band excites emission, which can be identify with transitions into photochromic centers. \textit{Ab initio} calculation of absorption spectrum of photochromic center agrees rather well with experimental data.
\end{abstract}


\end{frontmatter}


\section{Introduction}

Calcium fluoride crystals doped with certain rare earth ions (CaF$_{2}$:$RE$, $RE$=La$^{3+}$, Ce$^{3+}$, Gd$^{3+}$, Tb$^{3+}$, Lu$^{3+}$, Y$^{3+}$) demonstrate the photochromic behavior under x-ray or gamma irradiation and additive coloration. Photochromic centers are responsible for the intense absorption bands in visible wavelength range. On basis of an extensive study of optical and EPR behavior \citep{Anderson1970, Alig1971, Staebler1971, Bugaenko2008, Sizova2012} of this centers the two models of photochromic centers were proposed. 

In the first model photochromic effect occurs under thermally or optically stimulated electron transition from the divalent rare earth ion to the nearest neighboring anion vacancy. Thus, PC center is F-center disturbed by the nearest-neighbor trivalent rare earth ion, but ionized PC center (PC$^{+}$) is charged anion vacancy near the divalent rare earth ion \citep{Staebler1968, Bernhardt1971, Egranov2015}.

In the second and the widely accepted model thermally stable photochromic centers in CaF$_{2}$ crystals consist of one PC$^{+}$(RE) or two electrons PC(RE) bound at the anion vacancy adjacent to the trivalent rare earth ion RE \citep{Anderson1970, Alig1971}. Otherwise, PC(RE) center is F$'$-center, F-center having two electrons in the ground state, and PC$^{+}$(RE)-center is F-center disturbed by nearest-neighboring RE$^{3+}$ ion. This model is also confirmed by ENDOR for the Ce$^{3+}$ photochromic center in CaF$_{2}$~\citep{Aldous1976}. Colored crystals exhibit a photochromic effect, i.e. they change color under exposure to light. This process is accompanied by a reversible transformation of PC(RE) center \citep{Staebler1971, Bugaenko2008, Sizova2012}.
 
The data on the PC center luminescence could clarify the model of the photochromic center and explain the mechanism of its formation. However, no luminescence of F-like centers in alkaline earth fluorides has been observed yet. Furthermore, according to Bartram and Stoneham excited F-centers in fluoride crystals decay non-radiatively \citep{Bartram1975}. Nevertheless, \cite{Gorlich1968} and \cite{Kotitz1975} observed partially polarized broad band luminescence corresponding to PC(Y) and PC(La) centers in additively colored CaF$_{2}$-Y and CaF$_{2}$-La crystals. 

Attempts of theoretical calculations of photochromic centers were made to clarify mechanism of its formation and structure. \cite{Alig1971} showed in semiempirical calculation that bands in optical absorption spectra of PC$^{+}$(Ce) center were associated with transitions in a lowering of symmetry of F-center by neighbor rare earth ion. Overlap of 5d orbitals of rare earth ion with orbitals of F-center formed excited states of photochromic center.  \textit{Ab initio} calculations of PC(Y) and PC$^{+}$(Y) centers were performed by \cite{Mysovsky2008}. The authors calculated position of absorption bands of PC(Y) and PC$^{+}$(Y) centers. Unfortunately, the agreement of calculated optical absorption bands with experimental was worse than they expected. In the article by \citep{Mysovsky2011} optical absorption and luminescence spectra of perturbed F-center were calculated using  \textit{ab initio} method. Authors predicted luminescence of this centers in near infrared wavelength range. We employed a similar calculation method of hybrid embedded cluster.

In this article, we observe near infrared broadband luminescence associated with photochromic centers in the irradiated CaF$_{2}$ crystals doped with Lu ions. Our optical spectroscopy and theoretical calculation data give further information about electronic states of PC(Lu) centers.

\section{Methodology}
\subsection{Experimental technique}
Crystals of CaF$_{2}$ were grown from the melt by the Bridgman-Stockbarger method in graphite crucibles in vacuum and were doped with 0.1 mol.\% of LuF$_{3}$. In alkaline earth fluoride single crystal growth a small amount of CdF$_{2}$ was generally used as a scavenger in order to remove oxides contained in the raw materials. The crystals were irradiated at 300 K by x-rays from a Pd tube operating at 35 kV and 20 mA for one hour. After irradiation, the crystals acquired a golden-yellow color. Undoped crystals had no color after irradiation.

The optical absorption spectra were obtained on a Perkin-Elmer Lambda 950 UV/VIS/NIR spectrophotometer at 80 and 300 K at the Baikal Analytical Center for Collective Use, Siberian Branch, Russian Academy of Sciences. Photoluminescence (PL) measurements were conducted using a 700 W xenon arc lamp at 80 and 300 K in vacuum cold-finger cryostat. The spectra were detected with a MDR2 grating monochromator, a photomultiplier FEU-83 with Ag-O-Cs photocathode, and a photon-counter unit. The luminescence spectra were corrected for spectral response of detection channel. The photoluminescence excitation (PLE) spectra were measured with a grating monochromator MDR12 and Xe arc lamp. Excitation spectra were corrected for the varying intensity of exciting light due.

\subsection{Calculation details}

The \textit{ab initio} calculations were performed using a hybrid embedded cluster method which allowed to combine quantum mechanical (QM) cluster calculation and classically atoms described with shell model. The QM cluster with defect was surrounded by a large number ($\approx$700) of atoms which were described classically with the pair potentials. We used the pair potential parameters of Bukingem form which were the same as by \cite{Myasnikova2010}. About 50 cations between QM cluster and classical region were replaced with interface atoms with LANL1 ECP pseudopotentials specially optimized to minimize distortion of edge of QM cluster. All atoms of the QM, interface and classical region were allowed to relax during geometry optimization step. The described method allowed to represent the lattice distortion around defect with taking into account deformation and polarization of lattice. About 6000 fixed atoms surrounded classical region for the representation the correct Madelung potential inside classical region. For the density functional theory (DFT) calculations we used the modified B3LYP functional containing 40\% of Hartree-Fock and 60\% of DFT exchange energies which showed most adequate electron state localization and was successfully employed for DFT calculations of defects in fluoride crystals. Optical energies and dipole matrix elements of transitions were calculated with the time-dependent DFT (TD-DFT) method applied.

We used the GUESS computer code by \cite{Sushko2000} for geometry optimization step and Gaussian 2003 code \citep{g03} for TD DFT calculations. The applicability of the embedded cluster calculation method for point defects in ionic crystals was described by \cite{Mysovsky2011} and \cite{Myasnikova2012} in more details.

The calculations were performed in a cluster Ca$_{16}$F$_{33}$. The central fluorine atom was deleted for modeling vacancy and one nearest cation was replaced by lutetium ion. We used SDD basis setted on Lu$^{3+}$ ion and 6-311$^{+}$G$^{*}$ basis on fluorine and calcium ions. Moreover for correct representation of the F-center density we added the diffuse d-shell to calcium basis.  

\section{Results and discussion}

\begin{figure}[]
\centering
\includegraphics[width=0.5\textwidth]{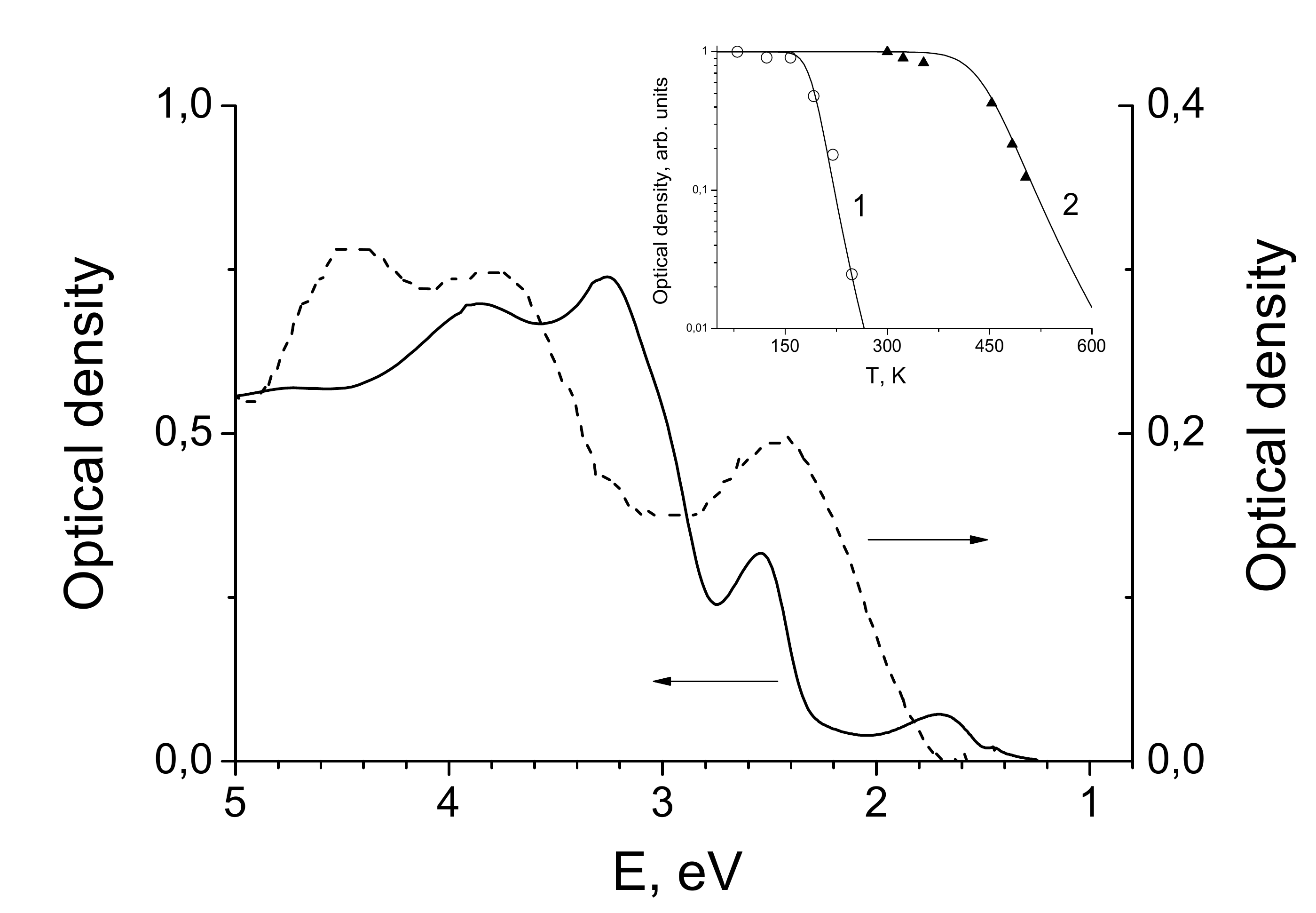}
\caption{Optical absorption spectra of CaF$_{2}$-Lu$^{3+}$ irradiated at room temperature (solid curve) and 80 K (dashed curve). In the inset temperature behavior of ionized photochromic center (curve 1) and photochromic center (curve 2) is shown }
\label{Fig1}
\end{figure}

If colorless CaF$_{2}$-Lu crystal is irradiated at room temperature it acquires golden-yellow. Its optical absorption spectrum is given in Fig.~\ref{Fig1}, solid curve. The color of crystals is due to an intense absorption in visible wavelength range. The absorption bands at 2.54; 3.25 and 3.83~eV are resolved. The bands do not resolved well at low temperatures down to 7.5~K. This bands are attributed to transition in photochromic PC(Lu) center. The ones are thermally bleached at about 600~K (see curve 2 of inset to the fig.~\ref{Fig1}).

The crystals irradiated at 80~K demonstrate other bands in optical absorption spectrum. The intense absorption bands at 2.44; 3.8 and 4.5~eV correspond to transition in ionized photochromic PC$^{+}$(Lu) center (Fig.~\ref{Fig1}, dashed curve). The spectrum of the PC(Lu) center at 80 K is identified by then photo-ionizing the aligned PC$^{+}$(Lu) center with visible (VIS) light and vise versa by photo-ionizing with ultraviolet (UV) light. Ionized photochromic centers become unstable at temperatures higher than 250~K (curve 1 of inset to the fig.~\ref{Fig1}). That is why only PC centers are observed in irradiated at room temperature crystals. Photochromic effect is observed only at low temperatures. The similar results were observed by \cite{Staebler1971} in the additively colored crystals.

We should also mention a band with energy 1.7~eV in the optical absorption spectra of PC(Lu) centers (see fig.~\ref{Fig1}, solid curve). In additively colored crystals in this band linear dichroism due to orientation of the PC(Lu) center was not observed by \cite{Staebler1971}. Also, this band is not involved in the reversible transformation of PC(Lu) center (photochromic effect) at low temperature. Therefore, we, as before \cite{Staebler1971}, do not attribute this band for transition into the PC(Lu) center.

\begin{figure}[]
\centering
\includegraphics[width=0.5\textwidth]{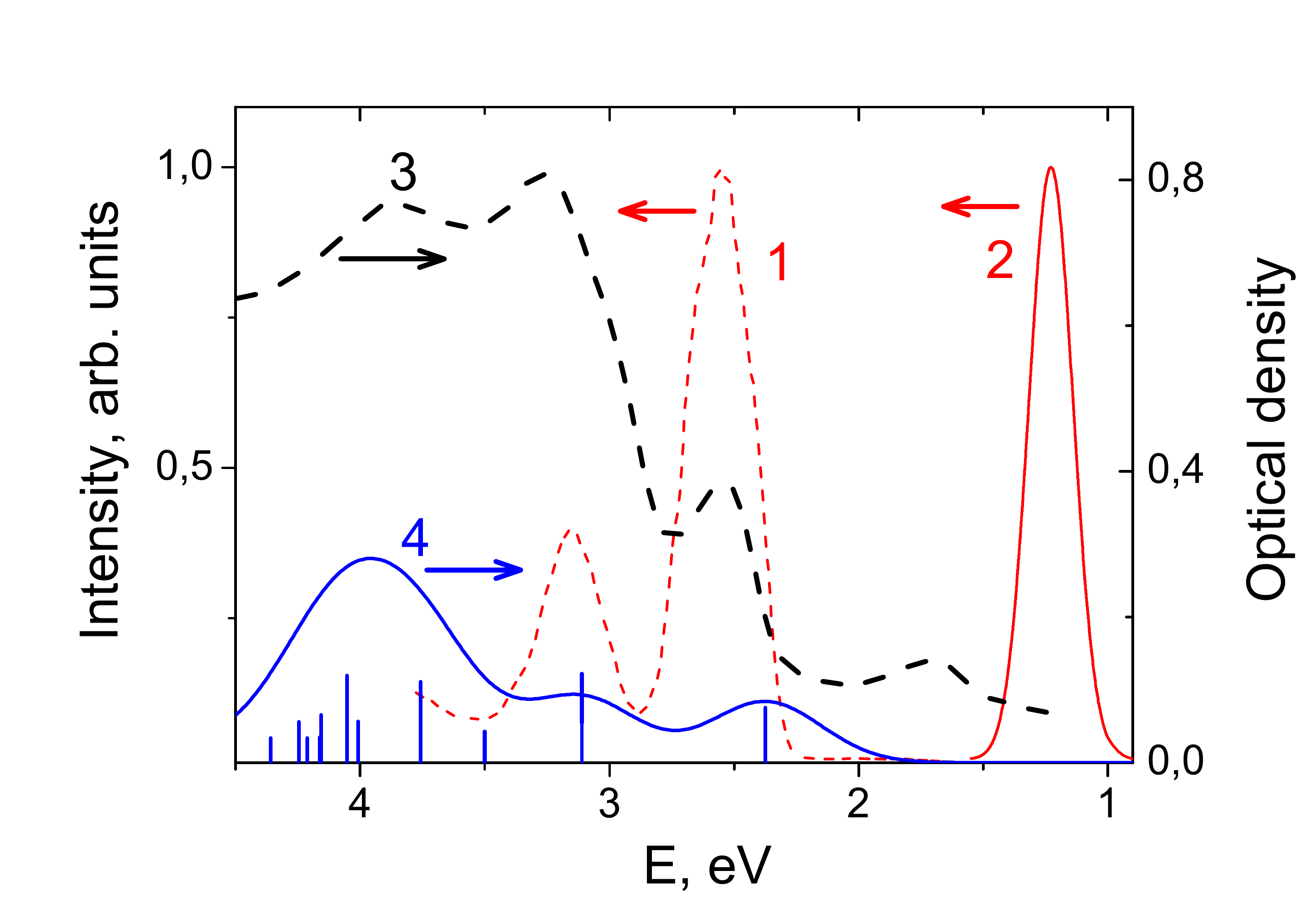}
\caption{Measured at 80~K excitation (curve 1), emission (curve 2), and optical absorption (curve 3) spectra of PC(Lu) centers in irradiated at 300~K crystals of CaF$_{2}$-Lu$^{3+}$. Vertical solid lines and curve 4 demonstrate calculated optical absorption spectrum of PC(Lu) center}
\label{Fig2}
\end{figure}

We find an intense luminescence in X-ray irradiated at 300~K CaF$_{2}$ crystals doped with Lu$^{3+}$ ions. The luminescence under lamp excitation in green wavelength range peaked at about 1.23 eV with a halfwidth of 0.21 eV (Fig.~\ref{Fig2}, solid curve 2). The excitation spectrum of the luminescence contains two bands peaking at 2.54 and 3.15 eV (Fig.~\ref{Fig2}, dashed curve 2). The low energy excitation band has more intensity than the higher energy one. The measured excitation spectrum also correlates well with the absorption spectrum of the PC(Lu) centers (Fig.~\ref{Fig2}, dashed curve 3). Therefore, we can conclude that the luminescence is due to radiative transitions into PC(Lu) center. 

\begin{figure}[]
\centering
\includegraphics[width=0.5\textwidth]{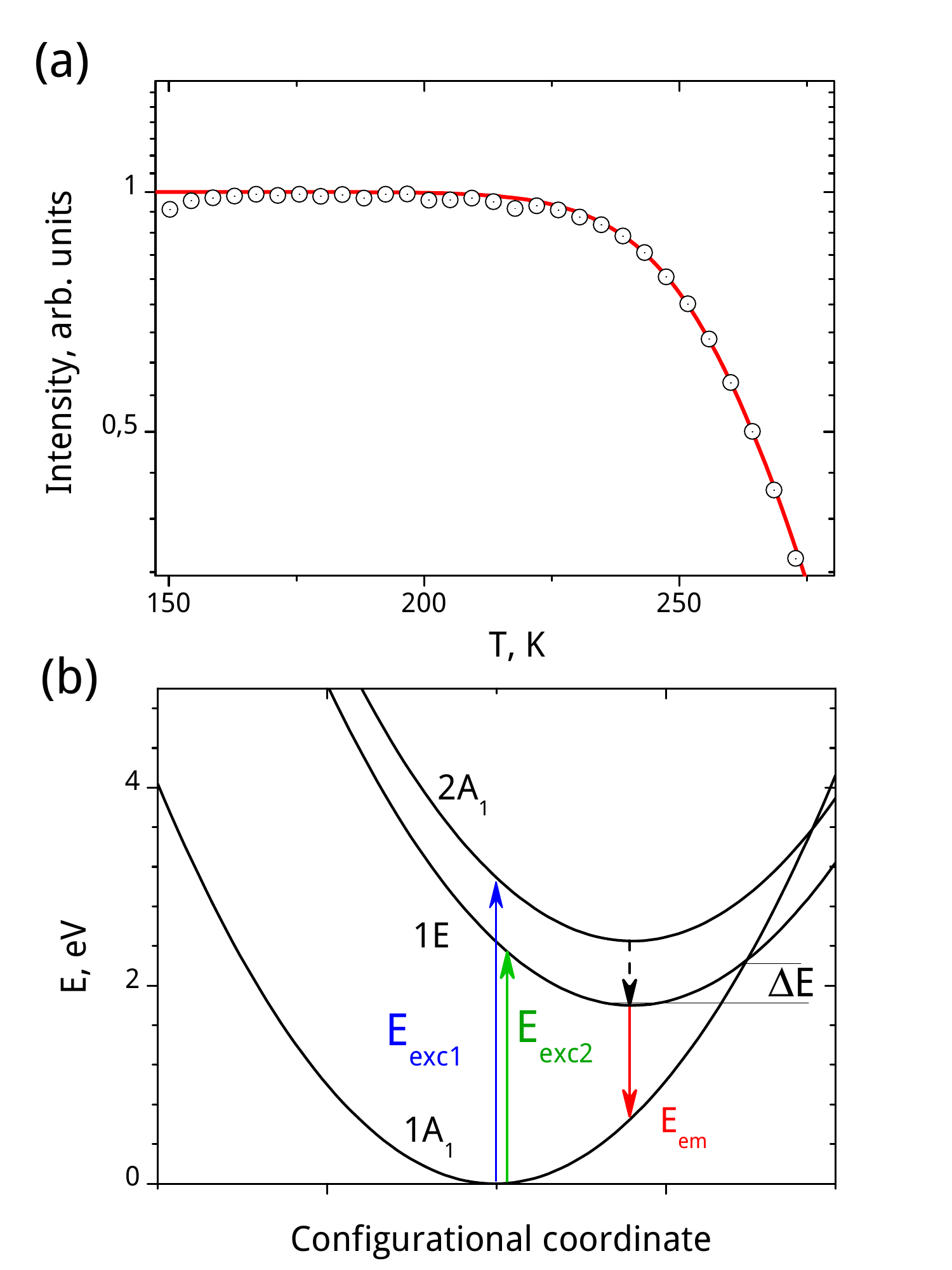}
\caption{Temperature dependence of luminescence intensity of PC(Lu) centers. Red curve is the Mott's equation fit (a). In subfigure (b) configuration coordinate model of absorption and emission processes associated with PC(Lu) centers is shown, here $E_{exc1}$=3.15~eV; $E_{exc2}$=2.54~eV; $E_{em}$=1.23~eV and $\Delta E$=0.47~eV estimated from eq.~(\ref{Mott_law})}
\label{Fig3}
\end{figure}

Intensity of the luminescence increases with decrease temperature (Fig.~\ref{Fig3}~(a)). The temperature dependence of the luminescence intensity is explained in terms of the probability of nonradiative transitions by Mott's equation \citep{Mott1938}:
\begin{equation}
\label{Mott_law}
I(T)=\frac{1}{1+w_{0}exp(-\Delta E/k_{B}T)},
\end{equation} 
with frequency factor $w_{0}=1.7\cdot10^{8}$. The activation barrier  for thermal quenching  can  be  estimated  at  the  value  of $\Delta E$=0.47~eV.

\begin{figure}[]
\centering
\includegraphics[width=0.5\textwidth]{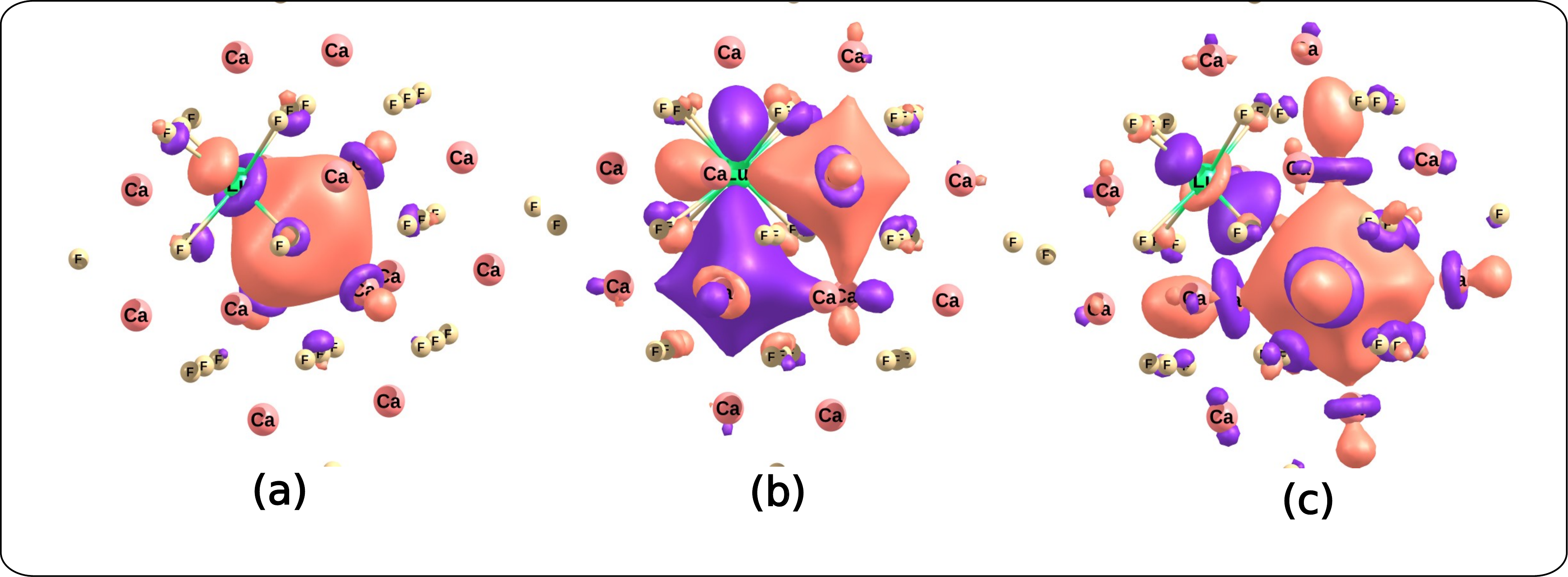}
\caption{Calculated one-electron states participating in the most intense optical absorption transition in PC(Lu) center}
\label{Fig4}
\end{figure}

The calculated lattice distortion of photochromic center is small. So in the fully relaxed configuration the three nearest cations displace inward about 0.04~\AA, and the displacements of six nearest cations do not exceed 0.11~\AA. However, lutetium ion is displaced about 0.13~\AA~to the fluorine vacancy direction from the starting position. The small deformation around F-center is quite typical for this defect as was mentioned by \cite{Mysovsky2011}. So the calculated structure of photochromic center  is correspond to F$'$-center having two electrons in ground state which is slightly disturbed by rare earth ion. One-electron ground state is shown in Fig.~\ref{Fig4}~(a). It is clear, that the ground level is a spin-singlet 1A state in which both electrons are in the lowest-energy spatial state (1s) with antiparallel spins.   

Excited state with energy 2.54~eV above the ground state is given in Fig.~\ref{Fig4}~(b). This state is formed by admixture of low-lying 5d states of rare earth ion and 2s-like states of F$'$-center. Higher energy states corresponding to transition with energy 3.25~eV is shown in Fig.~\ref{Fig4}~(c). It is formed by admixture of low-lying 5d states and 2p-like levels of F$'$-center. This result is in agreement with model proposed by \cite{Alig1971} for PC(Ce) center and results of calculation PC(Y) center reported by \cite{Mysovsky2008}.

In the calculated optical absorption spectrum three absorption bands are clearly observed (Fig.~\ref{Fig2}, vertical lines)). The smooth solid curve 4 shown in Fig.~\ref{Fig2} is a convolution of Gaussian-type functions, each centered at the excitation energy of the corresponding transition and weighed with the value of the corresponding oscillator strength. The full width at half maximum (FWHM) for all Gaussians is 1~eV. The band position is in rather well agreement with experimental data (within 0.1~eV).

Based on experimental and theoretical results we can construct the simple configurational coordinate diagram of luminescence process in PC(Lu) center (Fig.~\ref{Fig4}~(b)). In absorption of Lu-doped CaF$_{2}$, the allowed 1A$\to$1E and 1A$\to$2A transitions occur at 2.54 and 3.15~eV. Emission should then take place at longer wavelengths because of the expected Stokes shift. Thus, the 1.23 eV luminescence band is candidate for this emission. The absence of fine structure in the absorption (excitation) bands and strong temperature dependence of luminescence imply strong electron-phonon coupling for this transitions. Thermal activation energy of non-radiative intracenter process can be estimated about 0.47~eV. The relative weakness of the 3.15 eV band means a preferential population of the 1E level after excitation into optical absorption band of the PC(Lu) center.

\cite{Gorlich1968} observed polarized luminescence of PC(Y) center in Y$^{3+}$ doped CaF$_{2}$ and SrF$_{2}$ crystals. \cite{Kotitz1975} found weak luminescence in CaF$_{2}$ doped with lanthanum ions related to PC(La) centers. The luminescence was not detected at temperatures higher 40~K for the PC(Y) centers and higher 78~K for the PC(La) centers. That strong temperature dependence can be due to dominating non-radiative mechanism. PC center can be  considered as F-like center, therefore \cite{Bartram1975} estimation of the condition for luminescence to be observed is applied:

\begin{equation}
\label{bertr_law}
\Lambda'=(1-E_{emission}/E_{abs})/2\leq0.3.
\end{equation} 
Here $E_{emission}$ is energy for center of luminescence band and $E_{abs}$ is mean energy for optical absorption. 

For PC(Lu) centers we can calculate the ratio $\Lambda'$ from luminescence spectrum in Fig~\ref{Fig2}. It is approximately equal to 0.25. For PC(Y) and PC(La) centers using data of \cite{Gorlich1968} and \cite{Kotitz1975} the ratios are 0.4 and 0.28, respectively. 

The value of $\Lambda'$ is the lowest for PC(Lu) center and it demonstrates the brightest luminescence among all investigated PC centers due to lower probability of non-radiative transition. For PC(La) centers non-radiative recombination becomes more probable, however the thermal energy barrier $\Delta E$ is still high to observe luminescence at 78~K. In the case of PC(Y) centers the value of $\Lambda'$ is the largest, therefore the weak luminescence can be detected only at low temperature. PC(Ce) center has close to PC(La) energies for optical absorption bands, therefore, we can expect that PC(Ce) centers would demonstrate luminescence properties only at about 78~K. For other "photochromic" impurities the $\Lambda'$ value would be higher and the luminescence observes at lower than 77~K temperatures.   

\section{Conclusion}

We studied optical properties of photochromic centers in lutetium doped CaF$_{2}$ crystals. The photoluminescence in near IR wavelength range is attributed to photochromic centers. The theoretical calculation shows that excited states of PC(Lu) center have low-lying d orbitals of rare earth ion which overlap the F$'$-center wave functions. Therefore, the luminescence occurs due to transition from higher energy levels formed by admixture of p-like F$'$-center wavefunctions and low-lying d orbitals of rare earth ion to slightly disturbed 1s level of F$'$-center.

\section*{Acknowledgments}
The authors gratefully acknowledge A. V. Egranov for the fruitful discussions. The work was partially supported by RFBR grants 15-02-06666a and 15-02-06514a. The authors appreciate the use of Blackford computational cluster located at the Institute of System Dynamics and Control Theory SB RAS.



\bibliographystyle{elsarticle-harv} 
\bibliography{photochromic}

\end{document}